\documentclass[10pt,twocolumn,superscriptaddress,english,pra,showpacs,floatfix,aps]{revtex4-2}
\usepackage[utf8]{inputenc}
\usepackage{amsmath}
\usepackage{amssymb}
\usepackage{graphicx}
\usepackage{svg}
\usepackage{xspace}
    \usepackage{mathtools}

\usepackage{comment}
\usepackage{physics}
\usepackage{bm}

\usepackage{float}

\makeatletter
 
\usepackage[normalem]{ulem}
\usepackage{xcolor}
\usepackage{soul}
\usepackage[backref=none,
bookmarksnumbered=true,
bookmarks=true,
bookmarksopen=true,
colorlinks=true,
citecolor=blue,
linkcolor=blue,
anchorcolor=green,
urlcolor=blue,unicode=false]{hyperref}

\makeatother

\usepackage{babel}

\begin{document}

\title{Prethermal cooling with many-body quantum quenches}

\author{Jacob F.\ Steiner}
\affiliation{\mbox{Institute for Quantum Information and Matter,
California Institute of Technology, Pasadena, California 91125, USA}}

\author{Mohammad Hafezi}
\affiliation{Joint Quantum Institute, University of Maryland, College Park, Maryland 20742, USA}
\author{Stefan Kehrein}
\affiliation{\mbox{Institute for Theoretical Physics, Georg-August-Universität Göttingen, Friedrich-Hund-Platz 1, 37077 Göttingen, Germany}}

\author{Gil Refael}
\affiliation{\mbox{Institute for Quantum Information and Matter,
California Institute of Technology, Pasadena, California 91125, USA}}

\begin{abstract}
Many-body quantum quenches are typically associated with heating. In this work, we show that quantum quenches that perform positive work on the system can still lead to effective cooling of low-energy degrees of freedom if the quench energy is deposited in long-lived high-energy excitations. We discuss this explicitly for a quench of the hopping term~$t$ in the strong-coupling ($U \gg t$) fermionic Hubbard model at half filling, where the quench induces a very long-lived non-equilibrium doublon density. The associated prethermal state persists for a time exponentially large in $(U/t)^2$. During this time window, we find an effective prethermal temperature that is reduced by the square of the ratio of final to initial hopping amplitude with respect to the initial temperature. This manifests as an effective fluctuation-dissipation relation that holds for doublon-number conserving operators. In a practical implementation the Hubbard system acts as a refrigerant to cool a target system provided the coupling conserves doublon number. Our protocol can be thought of as a quantum quench many-body generalization of adiabatic demagnetization.
\end{abstract}


\maketitle 

\begin{figure*}[t]
    \centering
    \includegraphics[width=\linewidth]{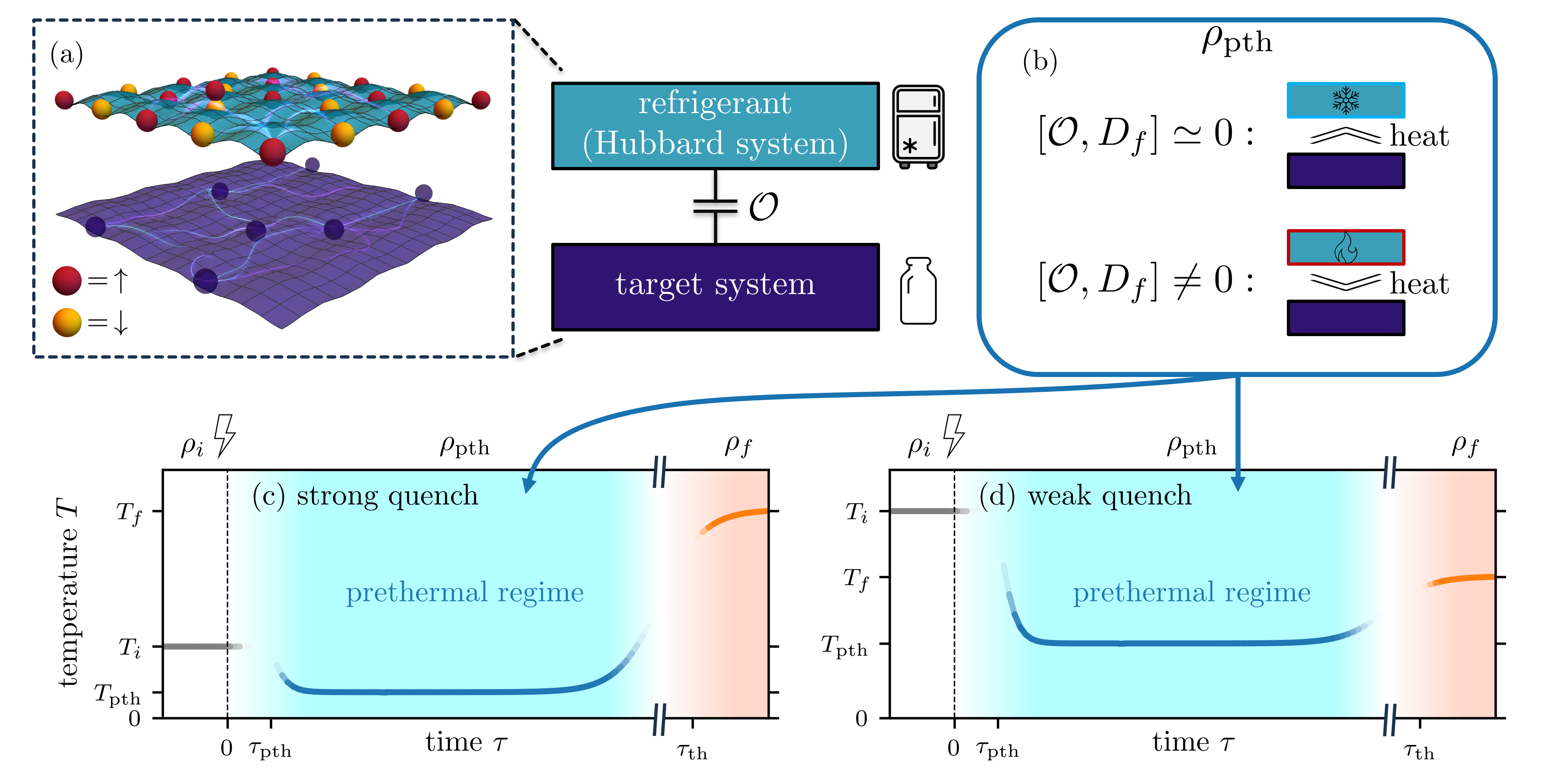}
    \caption{Prethermal cooling with quantum quenches.
    (a) Sketch of the setup. A Hubbard system, initialized in a thermal state $\rho_i$ at temperature $T_i$, is subjected to a hopping quench $t\to g_f t = t_f <t$ and coupled to a target system through operators $\mathcal O$.
    (b) For prethermal times $\tau_{\rm pth}\ll \tau \ll \tau_{\rm th}$, the doublon number is effectively conserved. If the coupling operator also conserves doublon number, $[\mathcal O,D_f]=0$, the target perceives the Hubbard system at the reduced temperature $T_{\rm pth}=g_f^2T_i$. Local density and spin operators are examples of such couplings. If $[\mathcal O,D_f]\neq0$, as for a local hopping operator, the target does not generally perceive a well-defined reduced temperature and may instead be heated.
    (c,d) Schematic time dependence of the effective temperature after a strong and weak quench, respectively. The system first relaxes on a time scale $\tau_{\rm pth}\sim J_f^{-1}$ to a prethermal state constrained by energy and doublon number. After the much longer time $\tau_{\rm th}\sim t_f^{-1}\exp{c(U/t_f)^2\ln[U/t_f]}$, the doublon number relaxes and the system reaches a thermal state at temperature $T_f$. For strong quenches, it is generically $T_f>T_i$, while for sufficiently weak quenches one may have $T_f\lesssim T_i$. 
    }
    \label{fig:fig_1}
\end{figure*}

\section{Introduction}
\label{sec:intro}
The creation of low-entropy quantum many-body states is essential for quantum simulation, and currently one of its key limitations 
\cite{bloch2008many,gross2017quantum,tarruell2018quantum}. Ultracold quantum gas experiments have realized the sub-nanokelvin regime using techniques like laser cooling and evaporative cooling \cite{Leanhardt2003}. However, the relevant quantity is not temperature but entropy per particle, which in state of the art experiments is still too high to observe the most exotic quantum phases. In particular, in the case of Fermi-Hubbard quantum simulators, a great experimental and theoretical effort has achieved regimes of
Mott physics, metal-insulator transitions, antiferromagnetic order
\cite{jordens2008mott,schneider2008metallic,hart2015observation,mazurenko2017coldatom,xu2025neutralatom},
and most recently signatures of stripe formation \cite{bourgund2025formation}, but the true low-energy regime where spin-liquid, pseudo-gap and high-temperature superconducting states  are expected \cite{lee2006doping, arovas2022hubbard} has so far been out of reach. 

In order to achieve even lower temperatures, a variety of new techniques have been put forward \cite{popp2006groundstate,griessner2006darkstate,capogrosso-sansone2008monte,ho2009squeezing,ho2009universal,deleo2011thermodynamics,paiva2011fermions,goto2017cooling,kantian2018dynamical,chiu2018quantum,mirasola2018cooling,werner2019lightinduced,werner2019entropycooled,zaletel2021preparation,kuzmin2022probing,wen2022floquets,langbehn2024dilute,matthies2024programmable,kishony2025efficiently,kishony2025gauged,langbehn2025universal,zhao2025feedback} and explored in ultracold atomic gases \cite{pinkse1997adiabatically,stamper-kurn1998reversible,bernier2009cooling,catani2009entropy,medley2011spin,chiu2018quantum,yang2020cooling,xu2025neutralatom}. One such route builds on the general idea underlying adiabatic demagnetization, where entropy is shifted from a target system to a refrigerant  \cite{stamper-kurn2009shifting}. In ultracold atomic gas experiments, the refrigerant can be either part of the gas itself \cite{bernier2009cooling,medley2011spin,chiu2018quantum,xu2025neutralatom}
or a second atomic species \cite{catani2009entropy,yang2020cooling}. 

The majority of protocols based on this general idea rely on adiabatic changes such that the target and refrigerant remain in equilibrium throughout the protocol \cite{pinkse1997adiabatically,stamper-kurn1998reversible,bernier2009cooling,catani2009entropy,ho2009squeezing,ho2009universal,medley2011spin,deleo2011thermodynamics,paiva2011fermions,chiu2018quantum,mirasola2018cooling,yang2020cooling,matthies2024programmable,kishony2025efficiently,kishony2025gauged,xu2025neutralatom}. While an adiabatic process gives the largest possible reduction in temperature, in many realistic situations limiting factors such as atom loss or coherence times demand finite preparation time. Hence, a number of works explore non-adiabatic cooling by quench-like changes of some control parameter \cite{goto2017cooling,kantian2018dynamical,werner2019lightinduced,werner2019entropycooled,zaletel2021preparation,wen2022floquets}.

Most of these non-adiabatic protocols rely on inhomogeneous quenches, where entropy is shifted between spatially separate parts of the atomic gas \cite{ho2009squeezing,ho2009universal,catani2009entropy,deleo2011thermodynamics,goto2017cooling,kantian2018dynamical,zaletel2021preparation,wen2022floquets,kuzmin2022probing}. For example \citet{zaletel2021preparation} discuss a cooling protocol where the overall scale of the Hamiltonian is suddenly reduced in part of the system, $H_i \to H_f = g H$ with $0 < g < 1$, trivially translating to a lower temperature in this region, $T_i \to T_f = g T$, as the state is unchanged by the quench, $\rho \propto e^{-H_i/T_i} = e^{-H_f/T_f}$. This lower temperature results in a heat flow from the remainder of the system (target region) into the quenched region (which acts as the refrigerant). In this manner one can generate a lower-entropy state in the target region. One major issue regarding the practical implementation of this protocol is whether one can experimentally change the overall energy scale of the Hamiltonian.

A very different non-adiabatic approach was put forward by \citet{werner2019lightinduced}: here, entropy is moved from a low-energy band near the Fermi level to a core band by means of a tailored spatially uniform, but temporally chirped pulse. The resulting state is very far from equilibrium due to the presence of holes in the core band. However, these holes are very long-lived, meaning the non-equilibrium state exhibits cold correlations in the low-energy band for a long prethermal time scale \cite{abanin2017rigorous}. 
Nonetheless, the chirp must be ramped up very slowly, essentially to give energy bins sufficient time to equilibrate, such that this protocol is again limited by its slowness.

In this work, we present a protocol that combines features of Refs.~\cite{zaletel2021preparation} and \cite{werner2019lightinduced}: namely a spatially uniform quench that leads to cooling over a prethermal time scale. Specifically, we consider a nontrivial quench of one parameter in a multi-parameter Hamiltonian, $H_i = H_0 + H_1 \to H_f = H_0 + g H_1$ with $0<g<1$ and $[H_0,H_1] \neq 0$. Such a quench has two opposing effects. First, it reduces energy scales, which tends to cool the system much like in the situation considered in Ref.~\cite{zaletel2021preparation}. Second, it creates high-energy excitations, which tend to increase the temperature once they equilibrate. Here, we focus on the nontrivial limit where the quenched term is subdominant, $\norm{H_0}\gg\norm{H_1}$, so that cooling is not simply inherited from a global reduction of the dominant energy scale.

The key point is that the excitations created by the quench can be extremely long-lived. Their lifetime defines a prethermal regime, characterized by the effective conservation of the number of high-energy excitations. Within this regime, the energy held by these excitations is not available to heat the low-energy sector, while the quench-induced reduction in energy scales still leads to an effective reduction in temperature. We show that this manifests as cold correlations through a fluctuation-dissipation relation satisfied by operators that conserve the number of high-energy excitations. We also find that, for strong quenches, the asymptotic final temperature reached after these excitations equilibrate is larger than the initial temperature, while for weak quenches, one can show using linear response that the final temperature may always be reduced albeit weakly if the quench parameter change is chosen with the correct sign. Thus, even when positive work is performed on the system, one can achieve effective cooling in a physically relevant part of the Hilbert space over prethermal times. Employing the quenched system as a refrigerant, and coupling it to a target system through operators that conserve the number of high-energy excitations, then suggests a route toward a fast, spatially uniform cooling protocol.

In the strong-coupling half-filled Hubbard model, a kinetic-energy quench realizes exactly this setting. Here, the long-lived excitations are doublons and holons \cite{winkler2006repulsively,strohmaier2010observation,sensarma2010lifetime,hassler2010dynamical,eckstein2011thermalization,kollar2011generalized,chudnovskiy2012doublon,hofmann2012doublon,abanin2017rigorous}, i.e. propagating doubly-occupied and empty sites. We henceforth specify to this situation, building on the established literature of quench dynamics in the Hubbard model \cite{moeckel2008interaction,eckstein2009thermalization,schiro2010timedependent}. Our protocol is illustrated in Fig.~\ref{fig:fig_1}. 

Our paper is structured as follows. In Sec.~\ref{sec:model} we introduce our model for quench cooling, namely the fermionic Hubbard model at half filling. Its quench dynamics is discussed in Sec.~\ref{sec:quench} and the ensuing prethermal cooling in Sec.~\ref{sec:prethermal_cooling}, which contains the main results of this paper. Section \ref{sec:thermal_cooling} deals with the asymptotic temperature beyond the prethermal regime, and Sec.~\ref{sec:discussion} closes with a discussion.  Technical details are delegated to appendices. 

\section{Model}
\label{sec:model}
To illustrate prethermal cooling under nontrivial quantum quenches, we consider the half-filled fermionic Hubbard model
\begin{equation}\label{eq:hamiltonian}
    H(g) = V + gK
\end{equation}
with interaction term 
\begin{equation}\label{eq:interaction_term}
    V = U \sum_j  n_{j,\uparrow}n_{j,\downarrow},\quad n_{j,\sigma} = c^\dagger_{j,\sigma} c^{}_{j,\sigma}
\end{equation}
and kinetic term
\begin{equation}\label{eq:kinetic_term}
     K = - t \sum_{\ev{jj'}}\sum_{\sigma} c^\dagger_{j,\sigma} c^{}_{j',\sigma}.
\end{equation}
The operator $c_{j,\sigma}$ annihilates ($c^\dagger$: creates) a spin-$\sigma$ fermion on site $j$, $n_{j,\sigma}$ is the associated occupation operator, and the sum over $\ev{jj'}$ is constrained to nearest neighbors $j$ and $j'$. The dimensionless parameter $g$ is introduced for later convenience. 

We focus on the strong coupling limit where the on-site repulsion $U$ dominates over the hopping amplitude $t$, $U \gg t$. In this regime, it is energetically favorable to have all sites singly occupied to avoid the interaction energy. Charge excitations, i.e. doubly occupied sites or ``doublons" (which at half-filling are accompanied by an equal number of empty sites or ``holons"), are thus subject to a gap $\sim U$ and the system forms an insulating state at temperatures $T \ll U$, the so-called Mott insulator. 
While charge excitations are frozen out at these temperatures, the remaining system of spin-$\frac{1}{2}$ degrees of freedom is much softer, governed by the Heisenberg exchange energy scale $J = 4t^2 / U \ll U$. In this work, we will consider temperatures $T \sim J$, where spin excitations are thermally populated while charge excitations remain frozen out. While we will consider a one-dimensional chain for numerical approaches below, the basic mechanism discussed here does not rely on one-dimensionality. Moreover, similar arguments should apply to strongly interacting bosons near unit filling, although we focus on fermions for definiteness.

The physics of $H(g)$ at temperatures $T \ll U$ is captured by an effective Hamiltonian that conserves the number of doublons. The effective Hamiltonian is obtained via Schrieffer-Wolff transformation \cite{macdonald1988expansion,bravyi2011schrieffer},
\begin{equation}
    \tilde{H}(g) = e^{S(g)} H(g) e^{S^\dagger(g)},
\end{equation}
where $S^\dagger(g) = -S(g)$ and 
\begin{equation}
    S(g) = \frac{g}{U} (K_+ - K_-) + \frac{g^2}{U^2} \bqty{K_+ + K_-,K_0} + \ldots.
\end{equation}
Here, we decomposed the kinetic energy operator $K = \sum_i K_i$ into terms $K_i$ that change the doublon number by $i = 0$, $+1$, or $-1$, respectively:
\begin{subequations}
\begin{align}
    K_0 =&\ - t \sum_{\ev{jj'}} \sum_\sigma \big( n_{j,\overline{\sigma}} c^\dagger_{j,\sigma} c_{j',\sigma}^{} n_{j',\overline{\sigma}} 
    \nonumber \\ 
    &\quad\quad\quad\quad\quad+ h_{j,\overline{\sigma}} c^\dagger_{j,\sigma} c_{j',\sigma}^{} h_{j',\overline{\sigma}}\big),\\
    K_+ =&\ - t \sum_{\ev{jj'}} \sum_\sigma n_{j,\overline{\sigma}} c^\dagger_{j,\sigma} c_{j',\sigma}^{}  h_{j',\overline{\sigma}}  ,\\
    K_- =&\ - t \sum_{\ev{jj'}} \sum_\sigma h_{j,\overline{\sigma}} c^\dagger_{j,\sigma} c_{j',\sigma}^{}  n_{j',\overline{\sigma}} ,  
\end{align}
\end{subequations}
where $h_{j,\sigma} = 1 - n_{j,\sigma}$ and $\overline{\sigma} = -\sigma$. 
We denote operators in the Schrieffer-Wolff frame by a tilde. To order $t^2/U$ the effective doublon-number-conserving Hamiltonian takes the form 
\begin{align}\label{eq:low_energy_ham}
    \tilde{H}(g) \simeq &\ V + g K_0 + \frac{g^2}{U} \bqty{K_+,K_-}.
\end{align}
The term $gK_0$ describes the motion of holons and doublons on the spin background. The last term contains several distinct processes. Acting on states without doublons or holons, it reduces to the effective antiferromagnetic Heisenberg Hamiltonian 
\begin{equation}\label{eq:heisenberg_ham}
    P_0\tilde{H}(g) P_0  = g^2 J\sum_{\ev{ij}} \pqty{\mathbf{S}_i\cdot\mathbf{S}_j - \frac{1}{4}}, 
\end{equation}
which describes the dynamics of the spin degrees of freedom in the Mott insulating state. Here, $P_0$ projects onto the $V = 0$ sector. Acting on states with doublons and holons, it contains correlated three-site hopping terms, which do not play an important role here and will be ignored in the following.

We note that not all double occupancies measured by $V$ in Eq.~\eqref{eq:hamiltonian} should be counted as physical doublon excitations. In particular, virtual double occupancies due to the Heisenberg exchange interaction do not propagate and hence should be excluded. To achieve this, note that by definition physical doublons are counted by $V/U$ in the Schrieffer-Wolff frame, as the transformation $e^{S(g)}$ is designed to precisely strip away the dressing with virtual double occupancies, and conversely, reintroduces it when transforming from the Schrieffer-Wolff to the bare frame. We thus define the density of doublon excitations in the Schrieffer-Wolff frame and the bare frame as
\begin{equation}
    \tilde{D} = V/U \textrm{ and } D(g) = e^{S^\dagger(g)}\tilde{D} e^{S(g)},
\end{equation}
respectively. Due to the dressing with virtual double occupancies encoded by $e^{S(g)}$ the doublon number operator $D(g)$ depends on relative magnitude of the interaction and kinetic terms, captured here by the dependence on the parameter $g$. To make the distinction between $V/U$ and $D(g)$ clear below, from now on we refer to the expectation value of $V/U$ in the bare frame as the number of double occupancies (as opposed to doublon number).

Before we proceed, we illustrate why this distinction is necessary with a simple example. One may show that the expectation value of the bare operator $V$ in a low-temperature state $T \sim J$ can be expressed in terms of the energy expectation value as $\ev{V} \simeq - \ev{H}$ 
\footnote{
This is easily seen by going to the Schrieffer-Wolff frame, $\tilde{V} = V + [S,V] + [S,[S,V]] + \ldots$, where the expectation value of $V$ and $[S,V]$ in a low-temperature state is exponentially suppressed in $U/T$. Only the term $P_0[S,[S,V]]P_0 = - P_0 \tilde{H}(g) P_0$ survives.
}. 
Noting that $\ev{H} < 0$ for such low-temperature states, it follows that lower energy states have a higher density of double occupancies. Of course this is not due to the presence of high-energy physical doublons, but due to the virtual double occupation involved in the Heisenberg exchange interaction. Therefore, $V/U$ is not the right operator to measure physical doublon density.  

While the effective doublon conserving model is a good description of the low energy physics, it obviously fails to capture doublon-holon recombination, see Sec.~\ref{sec:prethermal_cooling} below.  

\section{Quench and quench-induced doublon density}
\label{sec:quench}
We are interested in the change in temperature induced by a quantum quench of the kinetic term at time $\tau = 0$, 
\begin{equation}
    H_i \equiv H(g_i = 1) \to H_f \equiv H(g_f < 1)
\end{equation}
starting from the thermal state $\rho_i = \varrho(g_i = 1,T_i)$ at initial temperature $T_i$, and with energy $E_i$. Here, we defined the Gibbs state at temperature $T$ and parameter $g$, 
\begin{equation}
    \varrho(g,T) = \frac{ e^{-H(g)/T}}{\mathcal{Z}(g,T)},\quad \mathcal{Z}(g,T) =\tr\bqty{e^{-H(g)/T}}.
\end{equation}
The system then evolves according to
\begin{equation}\label{eq:rho_evolution}
    \rho(\tau) = e^{-i H_f \tau} \rho_i e^{i H_f \tau}.
\end{equation}

Consider the initial state with respect to the postquench Hamiltonian $H_f$. At temperatures $T_i \ll U$, $\rho_i$ has an exponentially suppressed doublon density $\ev{D_i}_i/N \simeq e^{-U/2T_i}/2$, where we defined $D_i = D(g_i)$ and $N$ is the total number of lattice sites. This suggests that one may use Eq.~\eqref{eq:heisenberg_ham} to describe the postquench system, seemingly implying a simple rescaling of the Heisenberg energy $J \to J_f = g_f^2 J$. By this logic, $\rho_i$ would also be a thermal state of $H_f$ (and thus stationary), with trivially rescaled energy $g_f^2 E_i$ and temperature $T_i \to g_f^2 T_i$. 

However, this is too naive: the quench changes the balance of kinetic and interaction energies, altering the dressing of the exchange interaction by virtual doublons. As the kinetic energy is reduced, $\rho_i$ contains too much virtual doublon dressing relative to $H_f$. The dressing due to exchange interactions is encoded in the transformation $e^{S(g_f)}$. Taking these observations into account one finds that the quench induces a non-equilibrium doublon number (see App.~\ref{app:postquench_state} for details)
\begin{equation}\label{eq:postquench_doublon_number}
    \ev{D_f}_i = - (1-g_f)^2 E_i / U, 
\end{equation}
and energy 
\begin{equation}\label{eq:postquench_energy}
    \ev{H_f}_i = (2g_f - 1) E_i.
\end{equation}
Here, we defined $D_f = D(g_f)$. Note that the induced doublon density is positive as the energy $E_i$ is negative at temperatures $T_i \sim J$. The approximation leading to Eqs.~\eqref{eq:postquench_doublon_number} and \eqref{eq:postquench_energy} clearly breaks down when $E_i > 0$ as negative doublon densities are unphysical. Furthermore, for $T_i \sim J$ it is $E_i \sim N J$ such that the quench-induced doublon density is of the order of $(t/U)^2 \ll 1$.

We assume that, after a sufficiently long time $\tau_\textrm{th}$, the system thermalizes under time-evolution. In dimensions $d>1$ this is expected provided the initial temperature is not too low. In $d=1$, the clean Hubbard model is integrable and technically does not thermalize; however, we assume that integrability is weakly broken by perturbations that do not qualitatively affect the strong-coupling physics. We do not explicitly model these perturbations. 

The thermalization time $\tau_\textrm{th}$ is set by the relaxation of the doublon density, which in the strong-coupling regime is exponentially slow \cite{strohmaier2010observation,sensarma2010lifetime,chudnovskiy2012doublon,abanin2017rigorous}. Other local observables relax on time scales set by the low-energy Heisenberg Hamiltonian. This separation of time scales defines a prethermal regime: after a prethermalization time $\tau_\textrm{pth} \sim J_f^{-1}$, the system reaches the prethermal state $\rho_\textrm{pth}$ with effective temperature $T_\textrm{pth}$. This state is studied in detail in Sec.~\ref{sec:prethermal_cooling}. 

The final temperature $T_f$ obtained in the long-time asymptotic regime is considered in Sec.~\ref{sec:thermal_cooling}. Interestingly, we find that even after complete thermalization, the temperature may be reduced for weak quenches, while for strong quenches it is increased compared to the initial temperature. 

The effective temperatures in the prethermal and thermal regimes and the associated separation of time scales are schematically depicted in Fig.~\ref{fig:fig_1}, where Fig.~\ref{fig:fig_1}(c) shows a strong quench with the asymptotic temperature increased compared to $T_i$, while Fig.~\ref{fig:fig_1}(d) shows a weak quench with reduced asymptotic temperature. In either case, the prethermal temperature is significantly smaller than $T_i$.

\section{Prethermal cooling}
\label{sec:prethermal_cooling}

This section analyzes the prethermal regime defined by the effective conservation of doublon number. We first analytically find the prethermal state by maximizing entropy subject to energy and doublon number conservation. Importantly, most of the quench energy enters as doublon density, where during prethermal times it is not available to drive an increase in temperature. This manifests as a strongly reduced prethermal temperature. To formalize this intuition, we introduce a target system and show via a fluctuation-dissipation relation that, if the coupling conserves doublon number, the target indeed perceives the Hubbard system at the reduced temperature $T_\textrm{pth}$. Finally, we corroborate this picture numerically. 

\subsection{Prethermal state}
A non-equilibrium doublon density relaxes via energy-conserving doublon-holon recombination accompanied by the emission of spin excitations. More specifically, a single energy-conserving doublon-holon recombination event requires the creation of on the order of $U/J_f \sim (U/t_f)^2$ spin excitations, corresponding to perturbation theory in $K$ to that order. This translates to a doublon decay rate \cite{strohmaier2010observation,sensarma2010lifetime,chudnovskiy2012doublon}
\begin{equation}
    \Gamma_d \sim t_f e^{-c\, (U/t_f)^2 \ln(U/t_f)},
\end{equation}
where $c > 1$ is a numerical constant. Here, we defined the shorthand $t_f = g_f t$. This is the longest time scale in the problem and therefore sets the thermalization time $\tau_\textrm{th} \sim 1/\Gamma_d$. For times $\tau \ll \tau_\textrm{th}$ the doublon number is effectively conserved. 

For prethermal times, i.e., times small compared to the thermalization time but large compared to the time scales at which other quantities relax ($\tau \gg  \tau_{\textrm{pth}} \sim 1/J_f$), the system instead relaxes to a state that maximizes entropy subject to conservation of energy and to the constraint that its doublon density matches the value directly after the quench \cite{kollar2011generalized,abanin2017rigorous}. This is achieved by the state
\begin{subequations} 
\begin{align}
    \rho_\textrm{pth} (T_\textrm{pth},\mu_\textrm{pth}) =&\ \frac{e^{-\pqty{H_f -\mu_\textrm{pth} D_f}/T_\textrm{pth} }}{\mathcal{Z}_\textrm{pth}(T_\textrm{pth},\mu_\textrm{pth})},\\
    \mathcal{Z}_\textrm{pth}(T_\textrm{pth},\mu_\textrm{pth}) =&\ \tr\bqty{ e^{-\pqty{H_f -\mu_\textrm{pth} D_f}/T_\textrm{pth} }},
\end{align}
\end{subequations}
where the Lagrange multipliers $T_\textrm{pth}$ and $\mu_\textrm{pth}$ are determined by the conditions (writing $\ev{\ldots}_\textrm{pth} = \tr[\ldots \rho_\textrm{pth}]$)
\begin{subequations} \label{eq:prethermal_eqs}
\begin{align}
    \ev{D_f}_i =&\ \ev{D_f}_\textrm{pth},\\
    \ev{H_f}_i =&\  \ev{H_f}_\textrm{pth}.
\end{align}
\end{subequations}
$T_\textrm{pth}$ acts as an effective temperature for the prethermal times but we emphasize that the state is not truly thermal. $\mu_\textrm{pth}$ acts as a chemical potential for doublons. Within the perturbative Schrieffer-Wolff framework and invoking that the induced doublon density is parametrically small, $\ev{D_f}_i/N \sim (t/U)^2$, we can approximately solve these equations (see App.~\ref{app:prethermodynamics}). We find for the doublon chemical potential
\begin{equation}
    \mu_\textrm{pth} \simeq  U -4d t_f + 2T_\textrm{pth}\ln\pqty{2\lambda^d \frac{\ev{D_f}_i}{N} }, 
\end{equation}
where $\lambda = \sqrt{4\pi t_f/ T_\textrm{pth}}$ is the de-Broglie wavelength of doublons and holons in this state. 
The prethermal temperature is to good approximation given by the naive scaling one would expect from the Heisenberg Hamiltonian,
\begin{equation}
    T_\textrm{pth} \simeq g_f^2 T_i. 
\end{equation}
This establishes an estimate of the parameters $T_\textrm{pth},\mu_\textrm{pth}$ which characterize the prethermal regime.

\subsection{Effective temperature from correlation functions}
We now ask the question to what extent the prethermal state can be used to cool a target system. 
To address this, we take the target system to be a two-level system (TLS) weakly and locally coupled to the Hubbard system (acting as refrigerant). The steady-state distribution of the TLS is controlled by the effective temperature. Explicitly, we consider 
$H \to H + H_\textrm{TLS}$ with
\begin{equation}
    H_\textrm{TLS} = \tfrac{1}{2} \delta E Z + \kappa X \mathcal{O},  
\end{equation}
where $Z,X,Y$ are Pauli operators of the TLS, $\delta E$ is the level splitting and $\kappa$ measures the coupling of the TLS to $\mathcal{O}$, a local operator of the Hubbard system. Assuming that $\kappa$ is sufficiently small and that the Hubbard system is in a stationary state, one may derive the Lindblad equation \cite{nathan2020universal} 
\begin{multline}\label{eq:lindblad_tls}
    \dot{\rho}_\textrm{TLS} = -i\frac{\delta E}{2} \bqty{Z,\rho_\textrm{TLS} } + \sum_{\pm} \bigg(L^{}_\pm \rho_\textrm{TLS}  L^\dagger_\pm \\ -\frac{1}{2} \Bqty{L^\dagger_\pm L^{}_\pm,\rho_\textrm{TLS} }\bigg).
\end{multline}
governing the dynamics of the TLS, with jump operators $L_\pm = \frac{1}{2}\sqrt{\gamma_\pm} (X\pm i Y )$ in terms of the rates 
\begin{equation}
    \gamma_\pm = \kappa^2 C (\mp \delta E).  
\end{equation}
Here, $C$ is the Fourier transform of the auto-correlation function of operator $\mathcal{O}$. $C$ can be written as
\begin{equation}\label{eq:correlator_0}
    C(\omega) = \int d \tau \, e^{i\omega  \tau} \ev{\mathcal{O}(\tau) \mathcal{O}(0)},
\end{equation}
where the expectation value is with respect to the stationary state. Evolving under Eq.~\eqref{eq:lindblad_tls}, the TLS approaches the state
\begin{equation}
    \rho_{\textrm{TLS}} \to \frac{1}{\gamma_+ + \gamma_-} \begin{pmatrix}
        \gamma_+ & \\ & \gamma_-
    \end{pmatrix}.
\end{equation}

For a thermal bath at temperature $T$ we expect this to reproduce the Boltzmann distribution. Specifically, the ratio between the excited and ground state populations should be $e^{-\delta E / T}$ for any choice of $\delta E$. This implies that $C$ needs to fulfill the detailed balance relation
\begin{equation}\label{eq:detailed_balance}
    \frac{\gamma_+}{\gamma_-} = \frac{C(-\omega)}{C(\omega)} = e^{-\omega /T},
\end{equation}
which obviously holds if the expectation value in Eq.~\eqref{eq:correlator_0} is with respect to a thermal state $\varrho(g,T)$. One can then read off the temperature of the bath by measuring the TLS distribution. In this sense the TLS acts as thermometer. 

Here, we are more interested in the effective temperature perceived by the TLS if the stationary state is taken as the prethermal state $\rho_\textrm{pth}$. One can readily show that the prethermal state satisfies the detailed balance relation (see App.~\ref{app:fdt_effective_temp} for details)
\begin{equation}\label{eq:fdt_pth}
    \frac{C_\textrm{pth}^{-\Delta D}(-\omega)}{C_\textrm{pth}^{\Delta D}(\omega)} = e^{-(\omega - \mu_\textrm{pth} \Delta D)/T_\textrm{pth}},
\end{equation}
where we introduced 
\begin{equation}\label{eq:correlator_pth}
    C_\textrm{pth}^{\Delta D}(\omega) = \int d\tau\, e^{i\omega\tau} \ev{\mathcal{O}^{-\Delta D}(\tau) \mathcal{O}^{\Delta D}(0)}_\textrm{pth},
\end{equation}
decomposing $\mathcal{O}$ into terms that change the doublon number by a fixed amount $\Delta D$, 
\begin{equation}
    \mathcal{O} = \sum_{\Delta D} \mathcal{O}^{\Delta D}.   
\end{equation}
Note that this expansion is defined with respect to the physical doublon number, i.e. it requires the Schrieffer-Wolff frame. However, operators that conserve the number of doubly occupied sites in the bare frame are dominated by the $\Delta D = 0$ term in the strong-coupling limit, $\mathcal{O} = \mathcal{O}^{\Delta D = 0} + \textrm{order}(t/U)$.

Equation \eqref{eq:fdt_pth} implies that a system that is coupled to the Hubbard system via a doublon-number conserving operator (i.e. with only the $\Delta D = 0$ term) indeed perceives the Hubbard system at the reduced temperature $T_\textrm{pth} = g_f^2 T_i$: in this case, $C_\textrm{pth}(\omega)  \simeq  C_\textrm{pth}^{\Delta D =0}(\omega)$, and therefore $C_\textrm{pth}(\omega)/C_\textrm{pth}(-\omega) = e^{-\omega /T_\textrm{pth}}$. Here, $C_\textrm{pth}(\omega)$ is given by Eq.~\eqref{eq:correlator_0} taking the expectation value with respect to $\rho_\textrm{pth}$. Conversely, if the coupling operators do not conserve doublon number, the induced distribution is highly non-thermal (i.e. the effective temperature extracted by the TLS depends on $\delta E$). 

\subsection{Numerical extraction of prethermal temperature}

\begin{figure*}[t]
    \centering
    \includegraphics[width=\linewidth]{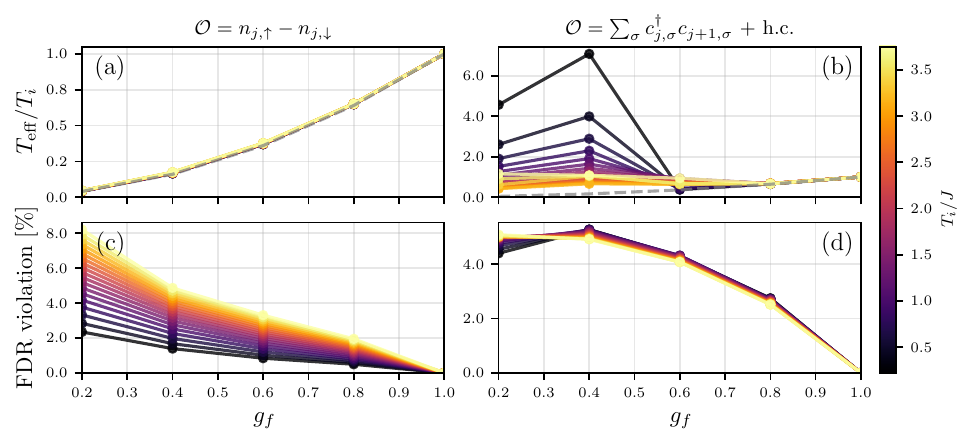}
    \caption{Numerical extraction of effective temperature from fluctuation-dissipation relation (FDR) for a one-dimensional Hubbard chain of length $N$, using the doublon number conserving coupling operator $\mathcal{O} =  n_{j,\uparrow} - n_{j,\downarrow}$ [panels (a,c)] as well as the doublon number nonconserving operator $\mathcal{O} = \sum_\sigma c^\dagger_{j,\sigma} c_{j+1,\sigma} + \textrm{h.c.}$ [panels (b,d)]. The first row [panels (a,b)] shows the extracted effective temperatures as a function of the quench parameter $g_f$ for different initial temperatures (color scale), the second row [panels (c,d)] shows the FDR violation defined in Eq.~\eqref{eq:fdt_violation} as a function of the same parameters. 
    Parameters: $U/t = 10$, $N = 10$ (see App.~\ref{app:fdt_more_data} for additional data and finite size scaling).}
    \label{fig:fig_2}
\end{figure*}

To corroborate these analytical findings, we numerically find the postquench correlation functions using exact diagonalization and extract an effective temperature using a fluctuation-dissipation criterion. To this end, we consider the postquench two-time correlation function
\begin{equation}\label{eq:correlator_1}
    C(\tau,\tau') =  \tr\bqty{\mathcal{O}(\tau)\mathcal{O}(\tau')\rho_i} = \tr\bqty{\mathcal{O}(\tau-\tau')\mathcal{O}\rho(\tau')},
\end{equation}
where time evolution is with respect to $H_f$. In particular, $\rho(\tau')$ is given by Eq.~\eqref{eq:rho_evolution}. $C(\tau,\tau')$ has the Lehmann representation
\begin{equation}\label{eq:correlator_2}
    C(\tau,\tau') = \sum_{lmn} \mathcal{O}_{lm} \mathcal{O}_{mn} [\rho(\tau')]_{nl} e^{i \varepsilon_{lm}(\tau-\tau')},
\end{equation}
where we introduced the matrix elements $\mathcal{O}_{lm} = \bra{\varepsilon_{l}} \mathcal{O}  \ket{\varepsilon_{m}}$ as well as $[\rho(\tau')]_{nl} = \bra{\varepsilon_{n}} \rho(\tau')  \ket{\varepsilon_{l}}$. Here, $\ket{\varepsilon_{m}}$ and $\varepsilon_{m}$ are eigenstates and energies of $H_f$, and we set $\varepsilon_{lm} =  \varepsilon_{l} -  \varepsilon_{m}$. Technically, due to the quench at time zero, Eq.~\eqref{eq:correlator_2} is the correct correlation function only for $\tau,\tau' >0$. However, we will take Eq.~\eqref{eq:correlator_2} as definition of the function $C_{AB}(\tau,\tau')$ for any value of $\tau$ \footnote{
This effectively removes energy non-conserving pieces which do not play a role at sufficiently late times.
}. This allows us to define the frequency space correlator via  
\begin{subequations}\label{eq:correlator_frq}
\begin{align}
     C(\omega;\tau') =&\  \int_{-\infty}^\infty d(\tau-\tau')\, e^{i\omega (\tau-\tau')}  C (\tau,\tau') \\
     \equiv&\ 2\pi \sum_{lm} \mathcal{M}_{lm}(\tau') \delta (\omega + \varepsilon_{lm}),
\end{align}
\end{subequations}
where we introduced the shorthand
\begin{equation}
     \mathcal{M}_{lm}(\tau') = \mathcal{O}_{lm} \sum_n \mathcal{O}_{mn} [\rho(\tau')]_{nl}. 
\end{equation}
Clearly, if $\rho(\tau')$ approaches a thermal state, $C(\omega;\tau')$ satisfies the detailed balance condition Eq.~\eqref{eq:detailed_balance}. Naively, one may thus want to define an effective temperature, by minimizing 
\begin{equation}
    \abs{C(-\omega;\tau')/C(\omega;\tau') - e^{-\omega/T}}
\end{equation}
with respect to $T$. However, this has two issues: first, for finite system sizes $C(\omega;\tau')$ does not have smooth support. Second, for small temperatures, the ratios in this expression can become very small. To bypass the first issue, we simply write the detailed balance condition for each transition separately: for a thermal state, this translates to $\mathcal{M}_{ml}/\mathcal{M}_{lm} = e^{-\varepsilon_{ml}/T}$. To alleviate the second issue, we introduce the fluctuation ('$+$') and dissipation ('$-$') functions $C^\pm(\omega;\tau') = \int_{-\infty}^\infty d(\tau-\tau')\, e^{i\omega (\tau-\tau')}  [C(\tau,\tau') \pm C(\tau',\tau)]$, which in equilibrium satisfy the fluctuation-dissipation relation $C^+(\omega;\tau')/C^-(\omega;\tau') = \coth\pqty{\omega /2T}$. This gives the condition $\mathcal{M}_{lm}^+ (\tau')/ \mathcal{M}_{lm}^- (\tau') = \coth\pqty{\varepsilon_{ml}/2T}$ for each transition $lm$, where we introduced
\begin{equation}
    \mathcal{M}_{lm}^\pm (\tau') = \mathcal{O}_{lm}  \bqty{ \mathcal{O} ,  \rho(\tau')}^\pm_{ml}, 
\end{equation}
where $\bqty{A ,  B}^\pm = AB \pm BA$. This motivates the definition of the effective temperature $T_\textrm{eff}$ through minimization of the function
\begin{equation}\label{eq:fdt_cost_function}
     f(T;\tau') = \norm{\mathcal{M}^- (\tau') - \mathcal{N}^+(T;\tau') }_F,
\end{equation}
with respect to $T$. Here, we introduced the shorthand $[\mathcal{N}^+(T;\tau')]_{lm} = \tanh\pqty{\varepsilon_{ml}/2T} \mathcal{M}_{lm}^+(\tau')$ and the Frobenius norm $\norm{A}_F = (\sum_{lm} \abs{A_{lm}}^2)^{1/2}$. Explicitly, the effective temperature is defined as
\begin{equation}
    T_\textrm{eff} = \underset{T}{\textrm{argmin}}\, f(T;\tau').
\end{equation}
This has the advantage over the naive previous approach that $\tanh\pqty{\varepsilon_{ml}/2T}$ nowhere becomes exponentially large or small. The normalized residual value 
\begin{equation}\label{eq:fdt_violation}
    \textrm{FDR violation}  = \frac{f(T_\textrm{eff};\tau')}{\norm{\mathcal{M}^- (\tau')}_F + \norm{\mathcal{N}^+(T_\textrm{eff};\tau') }_F }
\end{equation}
quantifies how strongly the fluctuation-dissipation relation is violated and thus measures the non-thermal character of the state. Note that it is $0 \leq \textrm{FDR violation} \leq 1$.

In essence, one may think about this approach as follows: generically, each transition $\varepsilon_{ml}$ favors a separate effective temperature obtained by minimization of each term in Eq.~\eqref{eq:fdt_cost_function} separately. This may be thought of as a frequency-dependent effective temperature $T_\textrm{eff}(\omega = \varepsilon_{ml})$. If $\rho(\tau')$ describes an equilibrium state at temperature $T$, all these effective temperatures agree and are identical to $T$. Away from equilibrium, this is not the case. It is still sensible to define an effective temperature if certain matrix elements are much larger than others, so that the sum in Eq.~\eqref{eq:fdt_cost_function} is dominated by terms which all give roughly the same effective temperature. The minimization attempts to identify this temperature. The FDR violation is a measure for the contribution from matrix elements that would favor a significantly different effective temperature. 

\subsection{Numerical results}

We now present the results of the numerical analysis based on the approach discussed in the previous section. To arrive at these results, we used exact diagonalization to find the eigenstates of $H_i$ and $H_f$ in a one-dimensional chain. With this knowledge we can evaluate $f(T;\tau')$ and minimize with respect to $T$. To avoid excessive memory costs, we truncate the spectra to the $n_\lambda$ lowest-energy states. To ensure that this does not discard important states, we bound the discarded weight in the initial state, and calculate the deviation from unitarity of the overlap matrix between eigenstates of $H_f$ and $H_i$. If either of those quantities exceeds a predefined threshold, we increase $n_\lambda$. For the system sizes considered, we find $n_\lambda = 5000$ to be sufficient (see App.~\ref{app:fdt_numerics} for details). We further observe that in practice, after a short initial transient, the results depend only weakly on $\tau'$. We interpret this as rapid relaxation of $\rho(\tau')$ to the prethermal state and identify the time scale of this relaxation as $\tau_\textrm{pth}$. To remove small fluctuations, we average the results over a few values of $\tau' \gg \tau_\textrm{pth}$.

Fig.~\ref{fig:fig_2} shows the resulting extracted temperatures (first row) and FDR violations (second row)  for a Hubbard chain of length $N=10$ with periodic boundary conditions. We consider correlations of the approximately doublon-number conserving $\mathcal{O} = n_{j,\uparrow} - n_{j,\downarrow}$ (first column) as well as of the doublon-number nonconserving $\mathcal{O} = \sum_\sigma c^\dagger_{j,\sigma} c_{j+1,\sigma} + \textrm{h.c.}$ (second column). By translation invariance, the results do not depend on $j$. In the doublon-number conserving case, the extracted temperature closely follows the prediction $T_\textrm{eff} = T_\textrm{pth} = g_f^2 T_i$ (dashed line) [panel (a)], independent of initial temperature. Conversely, in the doublon-number nonconserving case, the extracted effective temperature does not equal $T_\textrm{pth}$ for sufficiently strong quenches ($g_f \lesssim 0.6$) and becomes much bigger than the initial temperature for low initial temperatures [panel (b)]. In both cases, the FDR violation is in the few percent range and increases with decreasing $g_f$. Notably, in the doublon number conserving case, the FDR violation decreases as the initial temperature is decreased [panel (c)], while in the doublon number nonconserving case, it is essentially independent of initial temperature [panel (d)]. A systematic study of effective temperatures for different doublon-number nonconserving operators, and a detailed explanation of the behavior observed in panels (b), (c) and (d) go beyond the scope of this work.  

These results confirm our expectations discussed in the previous sections: the Hubbard system is perceived at the effective temperature $T_\textrm{pth}$ if the coupling is via a doublon-number conserving operator. In the doublon-number nonconserving case, the prethermal state is not perceived as thermal. 

Finally, we comment on the limitations of these numerical results. Due to the finite size gaps, the late time regime (i.e. the thermal limit) is not accessible. This is not a problem for the present case, as we are interested in the prethermal regime only. We interpret the convergence of the extracted effective temperatures with system size (see App.~\ref{app:fdt_more_data}) as evidence that the prethermal regime is well-described. 

\section{Thermal cooling}
\label{sec:thermal_cooling}

\begin{figure*}[t]
    \centering
    \includegraphics[width=\linewidth]{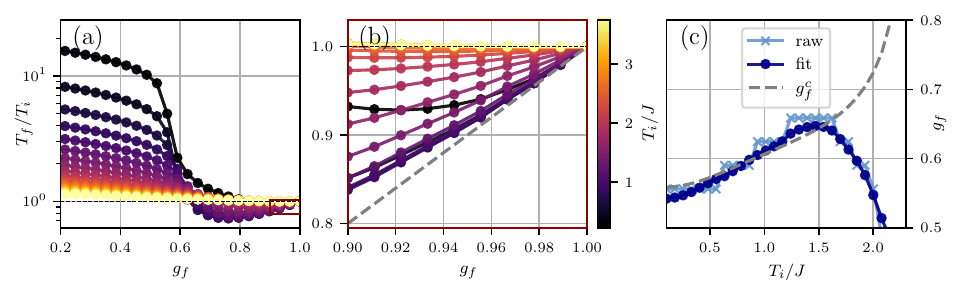}
    \caption{Asymptotic temperature after a quantum quench in the one-dimensional Hubbard model, obtained from the exact thermodynamic solution \cite{juttner1998hubbard}. Panel (a) shows a wide range of $g_f$ values, while panel (b) shows a zoom-in on the weak-quench regime near $g_f = 1.0$ together with the theoretical estimate from Eq.~\eqref{eq:linear_response_2}  (dashed gray line). The range shown in (b) is indicated by the red box in (a). The color scale in (a) and (b) denotes initial temperature, and is identical to that in Fig.~\ref{fig:fig_2}. 
    (c) Crossover point between the strong- and weak-quench regimes, determined from the maximum of $\abs{dT_f/dg_f}$. The light-blue data points are obtained from the raw numerical derivative of the data shown in (a), the dark-blue data points are obtained by a peak-fitting procedure that reduces the effect of the finite $g_f$-grid spacing. The dashed gray curve shows the estimate given in Eq.~\eqref{eq:crossover_g_f}. Parameters: $U=10.0t$.}
    \label{fig:fig_3}
\end{figure*}

Finally, we determine the asymptotic temperature $T_f$ reached after thermalization,
\begin{equation}
    \rho(\tau \gg \tau_\textrm{th}) \longrightarrow \rho_f = \varrho(g_f,T_f).
\end{equation}
$T_f$ is determined by conservation of energy alone, 
\begin{equation}\label{eq:thermal_equation}
    \ev{H_f}_i = \tr\bqty{H_f \rho_f} \equiv \ev{H_f}_f \equiv E_f.
\end{equation}
We can make progress by writing $\ev{H_f}_i = \ev{H_i}_i + (g_f - 1) \ev{K}_i$, or   
\begin{equation}
     E_f = E_i + (g_f - 1) [\partial_g F(g,T_i)]_{g=1}.
\end{equation}
Provided the thermodynamic functions $E(g,T) = \tr[H(g) \varrho(g,T)]$ and $F(g,T) = - T \ln \mathcal{Z}(g,T)$ are known, one may straightforwardly extract the final temperature from this equation.

In one spatial dimension the Hubbard model admits an exact solution through the Bethe Ansatz \cite{lieb1968absencea,mikeska1975correlation,juttner1998hubbard,deguchi2000thermodynamics,essler2005onedimensional}. We focus on this case, and employ techniques developed in Ref.~\cite{juttner1998hubbard} to find the curves $E(g,T)$ and $F(g,T)$ in the thermodynamic limit exactly. This is a novel application of Bethe Ansatz results. It permits us to obtain exact results about non-equilibrium processes for interacting systems in the thermodynamic limit, highlighting the importance of such exact methods. We stress that the fact that one needs to assume a small non-integrable term in order for the system to thermalize does not invalidate the applicability of the Bethe Ansatz: The weak integrability-breaking perturbation can be taken arbitrarily small, at the cost of an arbitrarily long thermalization time, so that its effect on the thermodynamic relation determining the asymptotic $T_f$ can be neglected.

The resulting values of $T_f$, shown as functions of $g_f$ for several initial temperatures, are plotted in Fig.~\ref{fig:fig_3}(a). We observe that for strong quenches ($g_f \ll 1$) the final temperature is increased compared to the initial temperature, especially for very low initial temperatures. Notably, for weak quenches ($1-g_f \ll 1$) the final temperature $T_f$ is appreciably reduced compared to $T_i$ for initial temperatures $T_i / J \lesssim 2.0$. Both the range of reduced temperature and the magnitude of the temperature reduction are maximal for $T_i \sim J$. 

The weak-quench cooling can be understood analytically from linear response. Solving Eq.~\eqref{eq:thermal_equation} to leading order in $0 < 1-g_f \ll 1$ gives (see App.~\ref{app:asymptotic_temp} for details)
\begin{equation}\label{eq:linear_response}
    \frac{T_f}{T_i} = 1 - (1- g_f) \frac{\ev{H_i K}_i - \ev{H_i}_i \ev{K}_i}{\ev{H_i^2}_i - \ev{H_i}_i^2}.
\end{equation}
In the low-temperature regime $T_i \sim J$, the numerator is positive and can be related to energy fluctuations of the effective Heisenberg model. Concretely, expanding to leading order in $t/U$, we find the relations
\begin{subequations}\label{eq:connected_correlator_HK}
\begin{align}
    \tr\bqty{\rho_i K} \simeq&\ 
    \frac{2}{U} \tr\bqty{\tilde{\rho}_i [K_+,K_-]} \simeq 2\ev{H_i}_i, \\
    \tr\bqty{\rho_i H_i K} \simeq&\ 
    \frac{2}{U^2} \tr\bqty{\tilde{\rho}_i [K_+,K_-]^2} \simeq 2\ev{H_i^2}_i, 
\end{align}
\end{subequations}
where we used that the physical doublon density is negligible in the initial state, such that one can restrict to the zero-doublon sector in the Schrieffer-Wolff frame, $\tilde{H}_i \to [K_+,K_-]/U$. We obtain
\begin{equation}\label{eq:linear_response_2}
    \frac{T_f}{T_i} = 2 g_f - 1.
\end{equation}
Indeed, for sufficiently small initial temperature, this behavior is reproduced by our numerical results, see the zoom-in in Fig.~\ref{fig:fig_3}(b). The dashed gray trace shows Eq.~\eqref{eq:linear_response_2}. For higher temperatures, the approximations in Eq.~\eqref{eq:connected_correlator_HK} break down. From Eq.~\eqref{eq:linear_response}, we still expect a reduction in temperature very close to $g_f = 1$ but this is not visible with our resolution in $g_f$. Conversely, as $T_i$ approaches zero, the deviation from the linear response curve occurs at weaker and weaker quenches, and the region of reduced temperature shrinks accordingly. This behavior is most clearly visible in the black curves in (a) and (b) with $T_i = 0.221 J$ (the lowest initial temperature shown in this figure). For $T_i = 0$, the temperature necessarily increases and our perturbative argument does not apply. Note that Eq.~\eqref{eq:linear_response_2} is consistent with the naive expectation $T_f/ T_i = g_f^2$ obtained by retaining only the Heisenberg Hamiltonian. This suggests that a weak quench does not create a sufficient number of excitations to invalidate the naive picture.

The strong-quench regime $g_f \ll 1$ is also analytically tractable. To linear order in $g_f$, one readily finds (see App.~\ref{app:asymptotic_temp} for details)
\begin{equation}
    T_f = \frac{U}{2\ell}  \pqty{1 - \frac{2}{\ell} g_f},
\end{equation}
where we introduced $\ell = \ln [U/(2\abs{E_i}/N)]$. This captures the large temperatures observed at small $g_f$. Here, the spin system can only store an energy $\sim J_f \ll J$ and the postquench energy translates directly to a physical doublon density. 

One can estimate the crossover point $g_f^c$ between the weak- and strong-quench regimes by asking where the ability of the spin system to absorb the full quench energy breaks down. Beyond this point, a thermal state is no longer possible without creating real doublon-holon excitations, implying a temperature $T_f \sim U$. The maximum energy stored by the spins at any positive temperature is reached at a temperature $T_f \gg J_f$ which is still small compared to the doublon energy, $T_f \ll U$. For such a temperature, one can replace the spin energy by its infinite temperature value, which, importantly, scales like $g_f^2$. This energy is $g_f^2 E_\textrm{s}^\infty$ with $E_\textrm{s}^\infty = - N z_\textrm{lat} J / 8$, where $z_\textrm{lat}$ ($= 2$ in $d=1$) is the lattice coordination. With this we find 
\begin{equation}\label{eq:crossover_g_f}
    g^c_f
    = \bqty{1+\sqrt{1 - \frac{E_{\textrm{s}}^\infty}{E_i}}}^{-1}. 
\end{equation}  
Under our assumptions, both $E_{\textrm{s}}^\infty$ and $E_i$ are negative and satisfy $0<E_{\textrm{s}}^\infty/E_i<1$. Therefore, the crossover lies in the range $1/2<g_f^c<1$ with the lower bound corresponding to low initial temperatures and the upper bound corresponding to high initial temperatures. For $d=1$, the minimum value of $g_f^c$ is reached if $E_i$ corresponds to the ground state energy $E_i = - N \ln(2) J$ such that $E_{\textrm{s}}^\infty/E_i= 1/(4\ln 2)$ and $g_f^c \simeq 0.56$. Fig.~\ref{fig:fig_3}(c) compares this estimate with the numerically obtained crossover point determined by the maximum rate of change of $T_f$ with respect to $g_f$. The agreement is surprisingly good for initial temperatures $T_i \lesssim J$ given the simplicity of this estimate.

\section{Discussion}
\label{sec:discussion}
We have shown that a quench of the hopping term in the half-filled strong-coupling Fermi-Hubbard model can lead to cooling over an exponentially long prethermal time scale provided that the temperature is probed by correlations of doublon-number conserving operators such as spin operators, see Fig.~\ref{fig:fig_2}. The Hubbard system can thus act as a refrigerant to cool a target system if the coupling operators conserve doublon number, see Fig.~\ref{fig:fig_1}. To run the protocol cyclically, the coupling is turned off after the target has equilibrated with the refrigerant, and the Hubbard refrigerant is reset to its prequench Hamiltonian and thermal state. This requires equilibration with a reservoir at temperature $T_i$. Then, the coupling may be turned on again, and the protocol is repeated. This is essentially the standard adiabatic demagnetization cooling cycle, only that the adiabatic ramp is replaced by a sudden quench. The quench protocol thus can be significantly faster than an adiabatic protocol if the equilibration steps require little time compared to the adiabatic ramp.

Intuitively, our protocol works because the work performed by the quench is mainly deposited in long-lived high-energy doublons and holons, while the low-energy Hilbert space of spin excitations is actually cooled due to the reduced Heisenberg exchange coupling. Doublon-number conservation violating processes connect the low and high-energy Hilbert spaces so that the system eventually thermalizes, however, this happens over an exponentially long time scale in the strong-coupling limit. Other protocols achieve cooling by a spatial separation between hot and cold regions \cite{ho2009squeezing,ho2009universal,catani2009entropy,deleo2011thermodynamics,goto2017cooling,kantian2018dynamical,zaletel2021preparation,wen2022floquets,kuzmin2022probing}, while our protocol achieves cooling by the effective separation of the total Hilbert space with respect to energy, or more specifically, doublon number.

Our protocol allows for practical implementation in ultracold atomic gas experiments, especially since spatially uniform quenches of the hopping terms can be realized easily by manipulating the lattice created by the counterpropagating lasers. In our numerical analysis we have focused on one-dimensional setups, but the quench cooling protocol straightforwardly carries over to higher dimensions. 

So far, we have focused on exact half filling. An interesting extension would be to study the doped case, where a finite density of holes is already present before the quench. Finally, another interesting direction is to further explore the numerical technique for effective temperature extraction based on the fluctuation-dissipation relation, in particular in the context of the eigenstate thermalization hypothesis. Related work in this direction includes Ref.~\cite{schonle2021eigenstate}.

\section{Acknowledgments}

We thank Andreas Klümper for providing his Bethe Ansatz code to solve the thermodynamics of the one-dimensional Fermi-Hubbard model used in generating the data in Fig.~\ref{fig:fig_3}. JFS and GR acknowledge the support of the AFOSR MURI program, under Agreement No. FA9550-22-1-0339. SK acknowledges support by the Deutsche Forschungsgemeinschaft (DFG, German Research Foundation) – 499180199; via FOR 5522, project T1.

\bibliographystyle{apsrev4-2}
\bibliography{library}

\appendix

%\clearpage
%\newpage 

\onecolumngrid

\section{Postquench expectation values in the prequench state}
\label{app:postquench_state}
The pre- and postquench generators of the Schrieffer-Wolff transformation are $S_i \equiv S(g=1)$ and $S_f \equiv S(g=g_f)$. In order to evaluate the expectation values of postquench observables in the prequench state within the low energy theory, we need to insert the product
\begin{equation}
    e^{S_i} e^{S_f^\dagger} \simeq e^{(1-g_f)S_i}.
\end{equation}
With this, we can express the postquench energy and physical doublon density through prequench expectation values. The postquench energy is
\begin{align}
    E_f =&\ \tr[H_f \rho_i]  = \tr[\tilde{\rho}_i  e^{S_i} e^{S_f^\dagger}  \tilde{H}_f e^{S_f}e^{S_i^\dagger}] \\
    \simeq &\ \tr\bqty{\tilde{\rho}_i  \Bqty{\tilde{H}_f + (1-g_f)\bqty{S_i,\tilde{H}_f}+ \frac{1}{2}(1-g_f)^2 \bqty{S_i,\bqty{S_i,\tilde{H}_f}} } }.
\end{align}
Using $\tilde{H}_f \simeq V + g_f K_0 + g_f^2 [K_+,K_-]/U$ and making use of the fact that the doublon density is exponentially suppressed in the initial state $\tilde{\rho}_i$ (at temperatures $T_i \sim J$), we reduce this to 
\begin{align}
    E_f =&\  (2g_f - 1)\tr\bqty{\tilde{\rho}_i [K_+,K_-]/U } .
\end{align}
The operator in the trace is proportional to the Heisenberg Hamiltonian. We can therefore express this through the prequench energy as
\begin{equation}
    E_f = (2g_f - 1) E_i.
\end{equation}
We can use similar considerations to find the postquench physical doublon density,
\begin{equation}
    \ev{D_f}_i = \tr\bqty{ D_f \rho_i } = \tr\bqty{ \tilde{\rho}_i  e^{S_i}e^{S_f^\dagger} \tilde{D} e^{S_f}e^{S_i^\dagger} } = \frac{1}{U}\tr\bqty{\tilde{\rho}_i \Bqty{V + (1-g_f) [S_i,V] + (1-g_f)^2 \frac{1}{2} [S_i,[S_i,V]]}}. 
\end{equation}
Under the same assumptions as before, this gives 
\begin{align}
    \ev{D_f}_i =&\ -(1-g_f)^2 E_i/U.
\end{align}

\section{Prethermodynamics}
\label{app:prethermodynamics}

Here, we find an approximate solution for Eqs.~\eqref{eq:prethermal_eqs}. It is advantageous to work in the low-energy picture, where
\begin{equation}
   \mathcal{Z}_\textrm{pth} = \tr e^{-\pqty{\tilde{H}_f - \mu_\textrm{pth} \tilde{D} }/T_\textrm{pth}} = \sum_{D} \mathcal{Z}(D) e^{z D},
\end{equation}
where we introduced $z = (\mu_\textrm{pth} - U)/T_\textrm{pth}$ as well as the partition sum at fixed number of $D$ doublons and $D$ holons,
\begin{equation}
    \mathcal{Z}(D) = \tr_{D} \bqty{\exp\Bqty{- \frac{1}{T_\textrm{pth}}\pqty{g_f K_0 + \frac{g_f^2}{U} \bqty{K_+,K_-}}}},
\end{equation} 
where the trace is over states with $D$ doublons and holons.

To make analytical progress, we make use of the fact that the postquench doublon and holon densities are of order $(t/U)^2 \ll 1$. We can thus neglect their interactions and quantum statistics, essentially reducing the problem to isolated doublons and holons moving on a correlated spin-background. Formally, this is done by a cluster-decomposition,
\begin{equation}
    -T_\textrm{pth} \ln [\mathcal{Z}(D)/ \mathcal{Z}_\textrm{s}] = 2D F_1 + F_\textrm{ent} + F_\textrm{int}, 
\end{equation}
where $\mathcal{Z}_\textrm{s} = \mathcal{Z}(0)$ describes the undisturbed spin system at the prethermal temperature, $F_1 = - T_\textrm{pth} \ln \mathcal{Z}_1$ is the free energy change associated with adding a single doublon or holon to the undisturbed spin system, $F_\textrm{ent} =  2T_\textrm{pth} \ln D!$ is the usual entropic contribution due to doublon (or holon) indistinguishability, and $F_\textrm{int}$ is the free energy contribution due to doublon-doublon, doublon-holon, and holon-holon interactions. The partition function of a single doublon or holon on top of the spin background is
\begin{equation}
    \mathcal{Z}_1 = \frac{1}{\mathcal{Z}_s}\tr_{1\textrm{d}} \bqty{\exp\Bqty{- \frac{1}{T_\textrm{pth}}\pqty{g_f K_0 + \frac{g_f^2}{U} \bqty{K_+,K_-}}}}. 
\end{equation}

The interaction contribution can be neglected if the typical interparticle distance is large compared to the spin correlation length. We assume this to be the case here, which should be warranted unless the temperatures are extremely low. To see that the doublon-holon gas is indeed in the nondegenerate regime, we use that $T_\textrm{pth} \lesssim J \ll t$ and anticipate that the doublon and holon chemical potentials sit at the bottom of the doublon band. Then, we may estimate the typical doublon and holon wavelength as (setting the lattice constant equal to one)
\begin{equation}
    \lambda = \sqrt{4\pi t_f/T_\textrm{pth}}. 
\end{equation}
We have to compare the associated volume $\lambda^d$ to the doublon and holon density,
\begin{equation}
    \lambda^d \frac{\ev{D_f}_i}{N} \sim (t_f/T_\textrm{pth})^{d/2} \frac{t_f^2}{U^2}.
\end{equation}
To be in the nondegenerate gas limit, we need this product to be small compared to $1$. For $T_\textrm{pth} \lesssim J$, this gives 
\begin{equation}
    \pqty{\frac{t_f}{U}}^{2 - \frac{d}{2}} \overset{!}{\ll} 1,
\end{equation}
which is satisfied in the strong coupling limit for any dimension $d < 4$. 

Using these approximations, we obtain  
\begin{equation} 
   \frac{\mathcal{Z}(D)}{\mathcal{Z}_\textrm{s}} \simeq \pqty{\frac{\mathcal{Z}_1^D}{D!}}^{2} \simeq \frac{1}{2\pi D}e^{-2D f_1/T_\textrm{pth}  + 2D \bqty{\ln (N/D) + 1}}, 
\end{equation}
where we introduced the free energy density of the single defect state, $f_1 = F_1 /N$. We can now find $\mathcal{Z}_\textrm{pth}$ using the saddle point method. Defining the doublon-holon partition function as $ \mathcal{Z}_\textrm{dh} = \mathcal{Z}_\textrm{pth}/\mathcal{Z}_\textrm{s}$, it is 
\begin{equation}
   \mathcal{Z}_\textrm{dh} \simeq \int_{-\infty}^\infty dD\,
   \frac{e^{D \bqty{\overline{z} + 2 \ln(N/D) +2}}}{2\pi D}  \simeq \frac{e^{2 D^*} }{2\sqrt{\pi D^*}}, 
\end{equation}
where we defined
\begin{align}
    \overline{z} =  z - 2 f_1/T_\textrm{pth} ,\quad  
    D^* = N e^{\frac{1}{2}  \overline{z} }.
\end{align}

We can now evaluate the expression for $\ev{D_f}_\textrm{pth}$ and $\ev{H_f}_\textrm{pth}$ and solve for the doublon chemical potential and prethermal temperature. We have $\ev{D_f}_\textrm{pth} = \partial_{z} \ln \mathcal{Z}_\textrm{pth}  \simeq D^*$. In order to satisfy the doublon number constraint $\ev{D_f}_\textrm{pth} = \ev{D_f}_i$, the chemical potential $\mu_\textrm{pth}$ needs to be 
\begin{equation}
    \mu_\textrm{pth} = U + 2 f_1 + 2T_\textrm{pth}\ln\pqty{\frac{\ev{D_f}_i}{N} },
\end{equation}
where $\ev{D_f}_i = -(1-g_f)^2 E_i/U$ as shown in App.~\ref{app:postquench_state}. Note that the last term is negative as $\lambda^d \ev{D_f}_i/N \ll 1$.
It remains to find the temperature from the constraint on energy, $\ev{H_f}_\textrm{pth} = \ev{H_f}_i$. One may express the prethermal temperature expectation value as 
\begin{equation}
    \ev{H_f}_\textrm{pth} = - \partial_{T_\textrm{pth}^{-1}} \ln \mathcal{Z}_\textrm{pth} + \mu_\textrm{pth} \ev{D_f}_\textrm{pth} \equiv E_\textrm{s} + E_\textrm{dh},
\end{equation}
where we decomposed the energy into spin and doublon-holon contributions, 
\begin{align}
    E_\textrm{s} =&\ - \partial_{T_\textrm{pth}^{-1}} \ln \mathcal{Z}_\textrm{s}, \\
    E_\textrm{dh} =&\ - \partial_{T_\textrm{pth}^{-1}} \ln \mathcal{Z}_\textrm{dh} + \mu_\textrm{pth} \ev{D_f}_\textrm{pth},
\end{align}
respectively. One readily finds
\begin{equation}
    E_\textrm{dh} \simeq \pqty{U + 2f_1 - 2 T_\textrm{pth} \partial_{T_\textrm{pth}} f_1 }D^* \simeq -(1-g_f)^2 E_i/U,
\end{equation}
where, noting that the leading terms in $f_1$ are of order $t_f$, in the second approximate equality we neglected terms of order $t^3/U^2$ and higher. Plugging this into the condition $E_\textrm{s} + E_\textrm{dh} \simeq (2g_f - 1) E_i$, we have 
\begin{equation}
    E_\textrm{s} \simeq g_f^2 E_i.
\end{equation}
The postquench spin system is described in terms of the Heisenberg Hamiltonian $H_\textrm{Hs}(g) = P_0\tilde{H}(g) P_0$, 
\begin{equation}
    \mathcal{Z}_\textrm{s} =  \tr\bqty{e^{- H_\textrm{Hs}(g_f)/ T_\textrm{pth}}}.
\end{equation}
Noting that $H_\textrm{Hs}(g_f) = g_f^2 H_\textrm{Hs}(g = 1)$ and that the prequench state has essentially no doublons, so that it is well described by the Heisenberg Hamiltonian $H_\textrm{Hs}(g = 1)$ at temperature $T_i$, we immediately obtain $T_{\textrm{pth}} \simeq g_f^2 T_i$. 

As a side note, we can obtain the thermal density of physical doublons by setting $g_f = 1$ and demanding that $\mu_\textrm{pth}$ vanish. We find
\begin{equation}
    \ev{D_i}_i \simeq N e^{-\frac{1}{2T_i} (U + 2 f_1)}.  
\end{equation}
As a crude estimate for $f_1$ can be obtained by treating the holon/doublon as a free particle on an uncorrelated spin background. This gives
\begin{equation}
    f_1 \simeq -2 d t  + T_i (d \ln \lambda + \ln 2).
\end{equation}
The second term in the bracket is due to the reduction in spin-entropy upon removal of a single spin. This gives 
\begin{equation}
    \ev{D_i}_i \simeq \frac{N}{2\lambda^d} e^{-\frac{1}{2T_i} (U-4dt)}.  
\end{equation}
In the estimate in the main text we ignore the contribution from hopping and simply set $\ev{D_i}_i \simeq N e^{-\frac{U}{2T_i}}/2$.

\section{Fluctuation-dissipation relation for the prethermal state}
\label{app:fdt_effective_temp}

Consider the prethermal state $\rho_\textrm{pth} \propto \exp{-(H_f - \mu_\textrm{pth} D_f)/T_\textrm{pth}}$. Here, we study to what extent this state satisfies a fluctuation dissipation relation. We will find that if the action of $\mathcal{O}$ conserves the doublon and holon numbers (as e.g. the low energy frame spin or density operators do), the corresponding correlation function satisfies the fluctuation-dissipation relation with effective  temperature $T_\textrm{pth}$ to very good approximation (corrections stem only from exponentially suppressed doublon non-conservation of $H_f$ itself). Conversely, if $\mathcal{O}$ changes doublon number, the correlations do not satisfy the usual fluctuation-dissipation relation, and the observed temperature depends on the energy - in other words the state is not thermal. We stress that the doublon number is defined with respect to the low energy frame. 

To get to these results, we make use of the Lehmann representation of the correlation functions $C_\textrm{pth}(\omega)$ defined in Eq.~\eqref{eq:correlator_0} taking the expectation value to be with respect to $\rho_\textrm{pth}$.  
Working in the Schrieffer-Wolff frame, and making use of the fact that doublon number can be treated as conserved in the prethermal regime, it is
\begin{equation}
    \tilde{\rho}_{\textrm{pth}} = \frac{1}{\mathcal{Z}_{\textrm{pth}}}\sum_{D} e^{\mu_\textrm{pth} D /T_\textrm{pth}} \sum_l e^{-\tilde{\varepsilon}_l^D/T_\textrm{pth}} \ketbra{\tilde{\varepsilon}_l^D}, 
\end{equation}
where $\ket{\tilde{\varepsilon}_l^D}$ are  eigenstates of $\tilde{H}_f$ with eigenenergy $\tilde{\varepsilon}_l^D$ in the doublon-holon sector $D$ (i.e. $\tilde{D} \ket{\tilde{\varepsilon}_l^D} = D \ket{\tilde{\varepsilon}_l^D}$). With this, one readily finds
\begin{equation}
    C_\textrm{pth}(\omega) = \frac{2\pi}{Z}\sum_{DD'}  \sum_{lm} \tilde{\mathcal{O}}^{DD'}_{lm}\tilde{\mathcal{O}}^{D'D}_{ml} \delta(\omega + \tilde{\varepsilon}_l^D-\tilde{\varepsilon}_m^{D'})e^{\mu_{\textrm{pth}} D/T_{\textrm{pth}} - \tilde{\varepsilon}_l^D/T_{\textrm{pth}}}. 
\end{equation}
Next, we expand the operator $\mathcal{O}$ into pieces that change the doublon number by a fixed amount. This is naturally done in the low energy frame,
\begin{equation}
    \tilde{\mathcal{O}} = \sum_{\Delta D} \tilde{\mathcal{O}}^{\Delta D},\quad [\tilde{D},\tilde{\mathcal{O}}^{\Delta D}] = \Delta D \tilde{\mathcal{O}}^{\Delta D}. 
\end{equation}
We also define the corresponding bare frame operators
\begin{equation}
    \mathcal{O}^{\Delta D} = e^{S^\dagger} \tilde{\mathcal{O}}^{\Delta D} e^S. 
\end{equation}
We obtain $C_{\textrm{pth}}(\omega) = \sum_{\Delta D} C_{\textrm{pth}}^{\Delta D} (\omega)$, where 
\begin{equation}
    C_{\textrm{pth}}^{\Delta D} (\omega) = \frac{2\pi}{Z}\sum_{D}  \sum_{lm} \tilde{\mathcal{O}}^{D,D+\Delta D}_{lm}\tilde{\mathcal{O}}^{D+\Delta D,D}_{ml} \delta(\omega + \tilde{\varepsilon}_l^D-\tilde{\varepsilon}_m^{D+\Delta D})e^{\mu_{\textrm{pth}} D/T_{\textrm{pth}} - \tilde{\varepsilon}_l^D/T_{\textrm{pth}}}. 
\end{equation}
It is straightforward to check that this satisfies Eq.~\eqref{eq:fdt_pth}, and, equivalently, that the correlation functions 
\begin{align}
    C_{\textrm{pth}}^{\Delta D;\pm}(\omega) =&\ C^{\Delta D}_{\textrm{pth}}(\omega) \pm C^{-\Delta D}_{\textrm{pth}}(-\omega) \\
     =&\  \frac{2\pi}{Z}\sum_{D} \sum_{lm}  
     \tilde{\mathcal{O}}^{D,D+\Delta D}_{lm}\tilde{\mathcal{O}}^{D+\Delta D,D}_{ml} \delta(\omega + \tilde{\varepsilon}_l^D-\tilde{\varepsilon}_m^{D+\Delta D})e^{\mu_{\textrm{pth}} D/T_{\textrm{pth}} - \tilde{\varepsilon}_l^D/T_{\textrm{pth}}}
     \pqty{1 \pm e^{ - (\omega - \mu_{\textrm{pth}}\Delta D)/T_{\textrm{pth}}  }}.
\end{align}
satisfy the fluctuation-dissipation relation
\begin{equation}
    \frac{C_{\textrm{pth}}^{\Delta D;+}(\omega)}{C_{\textrm{pth}}^{\Delta D;-}(\omega)} = \coth\pqty{\frac{\omega -\mu_{\textrm{pth}} \Delta D}{2T_{\textrm{pth}}}}.
\end{equation}
For doublon-number conserving $\mathcal{O}$ conserves doublon number only the term with $\Delta D = 0$ contributes, giving rise to the usual fluctuation-dissipation relation $C_{\textrm{pth}}^{+}(\omega)/C_{\textrm{pth}}^{-}(\omega) = \coth\pqty{\omega/2T_{\textrm{pth}}}$, where $C_{\textrm{pth}}^{\pm}(\omega) = \sum_{\Delta D}C_{\textrm{pth}}^{\Delta D;+}(\omega)$. This implies that a target system coupled to the Hubbard system (acting as refrigerant) through doublon conserving operators will perceive the prethermal state at the prethermal temperature $T_\textrm{pth}$, which due to the quench is reduced by the factor $g_f^2$ compared to the initial temperature. 

Finally, we discuss which operators conserve doublon number. Since this property is defined with respect to the low energy frame, operators that conserve the number of double occupancies in the bare frame will still have doublon-number nonconserving pieces. E.g., consider the operator that measures the number of double occupancies in the bare frame, $V/U$. In the Schrieffer-Wolff frame, this gives by construction $\tilde{V} \simeq V/U + [S,V/U] = V/U - (K_+ + K_-)/U$ to leading order in $t/U$. Thus, we have $\tilde{V}^{\Delta D = 0}/U \simeq V/U$ and $\tilde{V}^{\Delta D = \pm 1}/U \simeq K_\pm/U$. Transforming back, it is $V^{\Delta D = 0} \simeq V/U + K_+/U + K_-/U$ and $V^{\Delta D = \pm 1}/U \simeq K_\pm/U$ (as corrections are already of order $t^2/U^2$). Notably, for an operator that conserve the number of double occupancies in the bare frame, the doublon-number nonconserving terms are at least $t/U$ suppressed compared to the leading term. 
The situation is different for operators which do not conserve double occupancies already in the bare frame. For example consider the hopping operator on one link, 
\begin{equation}\label{eq:example}
    \mathcal{O} = \sum_\sigma c^\dagger_{j,\sigma}c_{j',\sigma} + \textrm{h.c.} = \mathcal{O}^1 + \mathcal{O}^0 + \mathcal{O}^{-1} + \order{t/U}.
\end{equation}
One can define $\mathcal{O}^1 + \mathcal{O}^0 + \mathcal{O}^{-1}$ in the same way one arrives at $K_i$ \cite{macdonald1988expansion}. This has doublon nonconserving terms already at order $1$. 
We thus say that double-occupancy conserving operators are \textit{approximately} doublon-number conserving. For simplicity, and since true doublon-number conservation would require fine-tuning of the coupling operators, in our numerical results, we contrast approximately doublon-number conserving operators and doublon-number nonconserving operators. 

\section{More details on the numerical scheme to extract an effective temperature from exact diagonalization data}
\label{app:fdt_numerics}

We numerically extract an effective temperature given an operator $\mathcal{O}$ and a state $\rho_i$ by minimizing the function $f$ defined in Eq.~\eqref{eq:fdt_cost_function} with respect to $T$. By definition, the minimum value of $f$ is attained at $T \equiv T_\textrm{eff}$ and we refer to this minimum value as FDR violation. To numerically evaluate $f$, we proceed as follows: 
\begin{itemize}
    \item We employ exact diagonalization techniques (exploiting conservation of charge) to find the lowest $n_\lambda$ states of the initial and final Hubbard Hamiltonians, $H_f$ and $H_i$, for a one-dimensional chain with $N$ sites at half-filling, i.e. the number of fermions is also $N$. Denote these eigenstates and energies as $\ket*{\varepsilon_{i,m}},\ \varepsilon_{i,m}$ and $\ket*{\varepsilon_{f,m}},\ \varepsilon_{f,m}$, respectively with $m \in \{1,\ldots n_\lambda\}$. 
    \item To ensure that sufficiently many eigenstates of $H_i$ are included, we check that the states up to $n_\lambda$ account for at least $99\%$ of the probability in the initial state. To put a bound on the discarded probabilities, we use
    \begin{equation}
         \textrm{neglected probability} = \frac{\sum_{m = n_\lambda + 1}^{\textrm{dim}(N)}  e^{-\varepsilon_{i,m}/T_i}}{\sum_{m=1}^{\textrm{dim}(N)} e^{-\varepsilon_{i,m}/T_i}} < [\textrm{dim}(N) - n_\lambda] \frac{e^{-\varepsilon_{i,n_\lambda}/T_i}}{ \sum_{m=1}^{n_\lambda} e^{-\varepsilon_{i,m}/T_i}}. 
    \end{equation} 
    Here, $\textrm{dim}(N)$ is the total Hilbert space dimension in the given symmetry sector for system size $N$. For $N=10$ and initial temperatures up to $T_i = 1.5 t$, we find $n_\lambda = 5000$ to be sufficient. 
    \item We calculate the $n_\lambda \times n_\lambda$ overlap matrices 
    \begin{equation}
         \mathcal{V}_{mn} = \braket*{\varepsilon_{f,m}}{\varepsilon_{i,n}}, 
    \end{equation}
    and the $n_\lambda \times n_\lambda$ matrices of observables of interest, 
    \begin{equation}
        \mathcal{O}_{mn} = \bra*{\varepsilon_{f,m}}\mathcal{O}\ket*{\varepsilon_{f,n}}. 
    \end{equation}
    \item To ensure that we have included sufficiently many eigenstates of $H_f$, we check that $\norm{\mathcal{V}^\dagger \mathcal{V} - I_{n_\lambda}}/n_\lambda < 0.01$, where $I_{n_\lambda}$ is the identity matrix of dimension $n_\lambda$. For $N=10$ and $n_\lambda = 5000$, we find this quantity to be a few $10^{-3}$. 
    \item The evolved state at time $\tau'$ is given by
    \begin{equation}
        [\rho_i(\tau')]_{nl} = e^{i\varepsilon_{f,nl}\tau'} \sum_p \mathcal{V}_{np} \frac{e^{-\varepsilon_{i,p}/T_i}}{\mathcal{Z}_i} \mathcal{V}^\dagger_{pl}.
    \end{equation}
    \item From this we find $\mathcal{M}^\pm_{lm} (\tau')$ and $f (T;\tau')$. We then minimize $f$ numerically with respect to $T$ using standard techniques.
\end{itemize}
We choose $\tau'$ sufficiently large such that convergence is achieved. We then average over $10$ values of $\tau'$. In the numerical results shown in this work, these values are $t_i\tau' \in \{52.63, 57.89, 63.16, 68.42, 73.68, 78.95, 84.21, 89.47, 94.74, 100.00\}$. 

\section{Additional data}
\label{app:fdt_more_data}
Here, we provide additional data obtained by the fluctuation-dissipation minimization algorithm described above. 

\subsection{Weaker interactions}

\begin{figure}[H]
    \centering
    \includegraphics[width=0.8\linewidth]{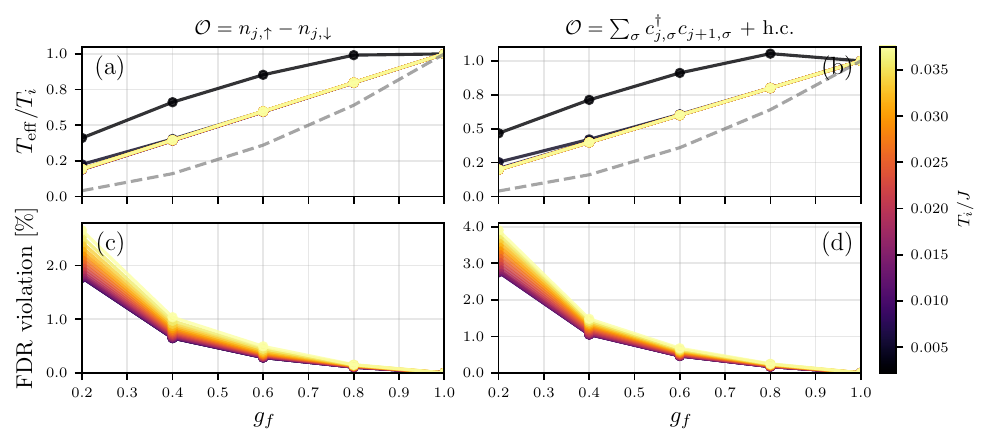}
    \caption{ Parameters: $U/t = 0.1$, $N=10$.}
    \label{fig:fdt_app_0.1}
\end{figure}

\begin{figure}[H]
    \centering
    \includegraphics[width=0.8\linewidth]{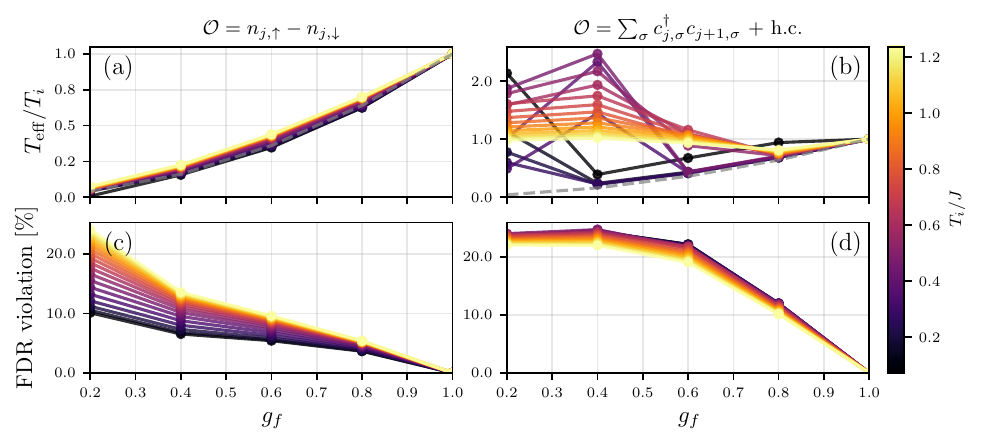}
    \caption{Parameters: $U/t = 3.3$, $N=10$.}
    \label{fig:fdt_app_3.3}
\end{figure}

Fig.~\ref{fig:fdt_app_0.1} shows data for very weak interactions, $U = 0.1t$. The behavior is essentially that of an overall quench of the Hamiltonian $H  \to g_f H$, leading to a trivial linear temperature reduction regardless of coupling operator. Note that the lowest initial temperature is comparable to the finite size gap (black data points). We do not expect the approach to work well in this extreme limit. 
Fig.~\ref{fig:fdt_app_3.3} shows data for $U = 3.3t$ which in essence already behaves as the strong coupling case considered in Fig.~\ref{fig:fig_2} in the main text.

\subsection{\texorpdfstring{Finer $g_f$-grid for smaller system size}{Finer gf-grid for smaller system size}}
\begin{figure}[H]
    \centering
    \includegraphics[width=0.8\linewidth]{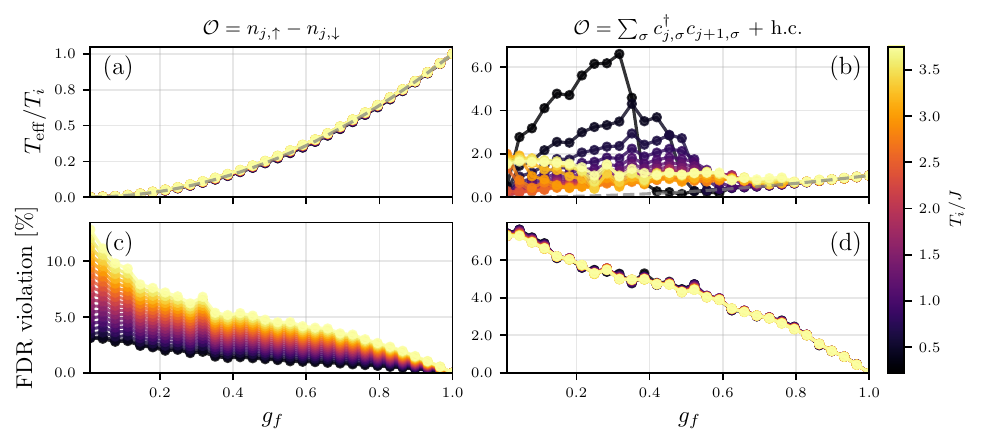}
    \caption{Parameters: $U/t = 10.0$, $N=6$.}
    \label{fig:fdt_app_finer_grid}
\end{figure}

Here, we show data on a finer grid of $g_f$-points to make sure that we do not miss essential features. As each $g_f$-point requires diagonalization of the Hamiltonian, we here consider only the smaller system size $N=6$. The resulting data is shown in Fig.~\ref{fig:fdt_app_finer_grid}.  

\subsection{Finite-size scaling}

\begin{figure}[H]
    \centering
    \includegraphics[width=0.8\linewidth]{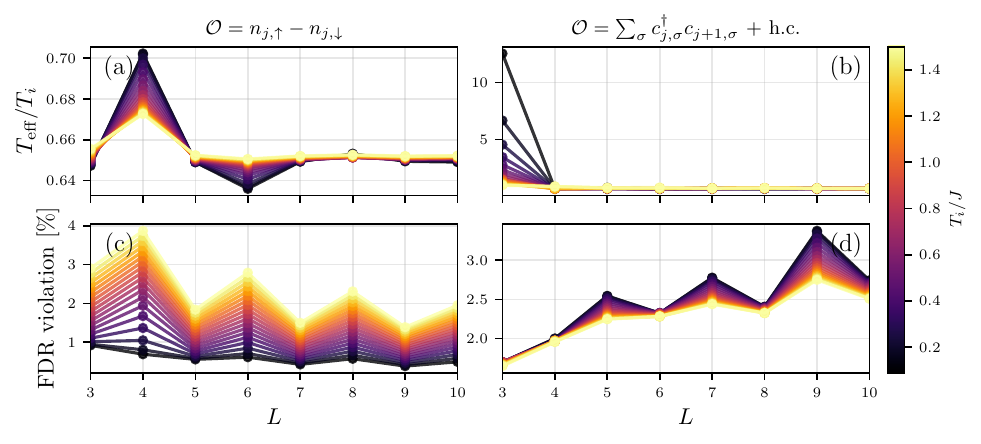}
    \caption{Parameters: $U/t = 10.0$, $g_f = 0.8$.}
    \label{fig:fdt_app_finite_size_0.8}
\end{figure}

\begin{figure}[H]
    \centering
    \includegraphics[width=0.8\linewidth]{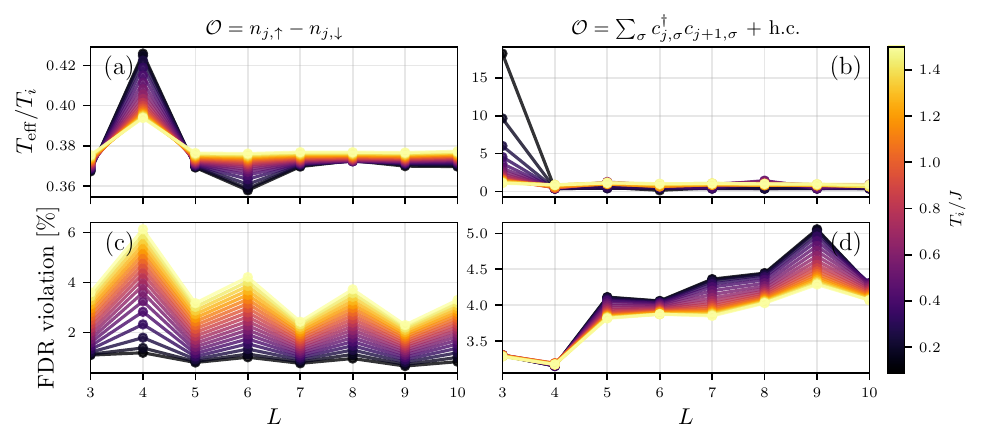}
    \caption{Parameters: $U/t = 10.0$, $g_f = 0.6$.}
    \label{fig:fdt_app_finite_size_0.6}
\end{figure}

\begin{figure}[H]
    \centering
    \includegraphics[width=0.8\linewidth]{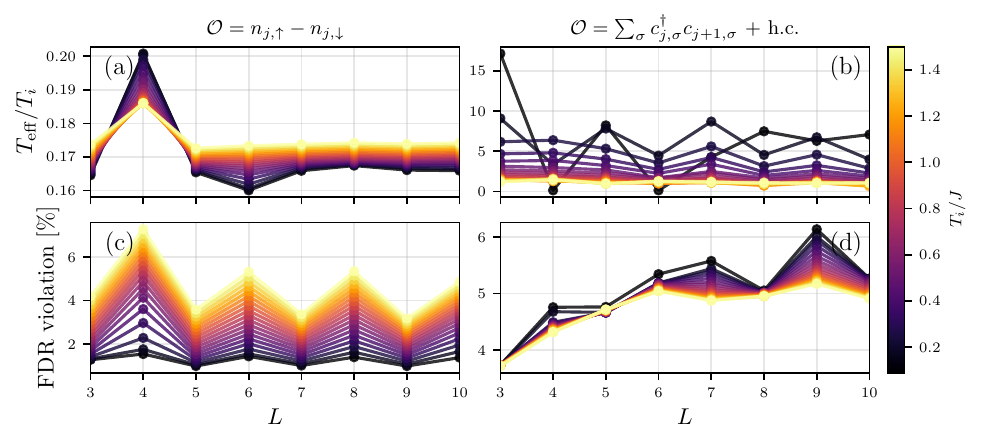}
    \caption{Parameters: $U/t = 10.0$, $g_f = 0.4$.}
    \label{fig:fdt_app_finite_size_0.4}
\end{figure}

\begin{figure}[H]
    \centering
    \includegraphics[width=0.8\linewidth]{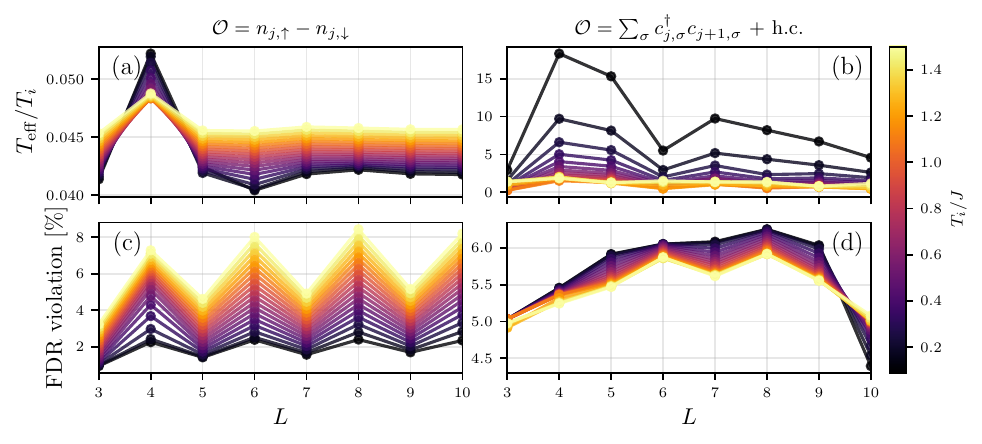}
    \caption{Parameters: $U/t = 10.0$, $g_f = 0.2$.}
    \label{fig:fdt_app_finite_size_0.2}
\end{figure}

Finally, we check convergence of the extracted temperature with system size, see Figs.~\ref{fig:fdt_app_finite_size_0.8} through \ref{fig:fdt_app_finite_size_0.2}. We note an even-odd effect in system size for the FDR violation, but not for the extracted temperatures which appear well-converged. 


\section{Analytical estimates for the asymptotic temperature \texorpdfstring{$T_f$}{Tf}}
\label{app:asymptotic_temp}

One can find approximate expressions for the energy and temperature in the final state using similar arguments as were used to describe the prethermal state. To this end, one again argues that it is possible to approximately decompose
\begin{equation}
    \mathcal{Z}_f = \tr\bqty{e^{-H_f/T_f}} \simeq \mathcal{Z}_\textrm{s} \mathcal{Z}_\textrm{dh},
\end{equation}
where $\mathcal{Z}_\textrm{s} = \tr\bqty{e^{-H_\textrm{Hs}(g_f)/T_f}}$ describes the spin system without any doublon or holons. The doublon-holon partition may be obtained by repeating the steps in App.~\ref{app:prethermodynamics} but without doublon chemical potential, giving
\begin{equation}
     \ln \mathcal{Z}_\textrm{dh} \simeq 2N e^{-\frac{1}{2T_f} (U+ 2 f_1)}.
\end{equation}
Recall that $f_1$ is the change in free energy density due to a single delocalized doublon or holon. As the final temperature may be large compared to the bandwidth $t_f$, it is no longer sufficient to expand the doublon dispersion around the band bottom. We therefore use the more general expression $f_1 = T_f [\ln 2 - d \ln I_0(2t_f/T_f)]$, where $I_n$ is a modified Bessel function of the first kind of order $n$. 
This expression is obtained by treating the doublon or holon as a particle in a cosine band on top of an uncorrelated spin background,
\begin{equation}
    f_1 \simeq T_f \ln 2 - T_f \ln \int \frac{d^d k}{(2\pi)^d} e^{(2t_f/T_f) \sum_{i=1}^d \cos k_i}. 
\end{equation}
The first term stems from the reduction in spin entropy upon removing one spin. While this expression assumes a square lattice, we do not expect the results below to be qualitatively different in other lattices.
The doublon holon energy is then
\begin{equation}
    E_\textrm{dh}(g_f,T_f) = N \bqty{\frac{U}{2} - 2d t_f r(2t_f/T_f)} [I_0(2t_f/T_f)]^d e^{-\frac{U}{2T_f}},\quad r(x) = \frac{I_1(x)}{I_0(x)}.
\end{equation}
Energy conservation then demands
\begin{equation}
    E_\textrm{s}(g_f,T_f) + E_\textrm{dh}(g_f,T_f) \simeq (2g_f - 1) E_i,
\end{equation}
where $E_\textrm{s}(g_f,T_f)$ is the energy of the Heisenberg system described by $H_\textrm{Hs}(g_f) = P_0 \tilde{H}(g_f) P_0$. 

Consider first weak quenches, $1-g_f \ll 1$. For such quenches, one can show for a general Hamiltonian quench $H_i \to H_f = H + \delta H$ that the temperature change $\delta T = T_f-T_i$ satisfies to leading order
\begin{equation}
    \frac{\delta T}{T_i} =  \frac{\ev{H_i \delta H}_i - \ev{H_i}_i \ev{\delta H}_i}{\ev{H_i^2}_i - \ev{H_i}_i^2} \equiv  \frac{\textrm{Cov}_i\pqty{H_i,\delta H} }{\textrm{Var}_i\pqty{H_i}}.
\end{equation}
In our setup, $\delta H = (g_f - 1) K$ such that the temperature change $\delta T$ is negative for positive covariance between $H_i$ and $K$. This is discussed in detail in the main text.

Conversely, for strong quenches normally no expansion is available. However, in our specific case, the point $g_f = 0$ is again very simple such that one can also do straightforward perturbation theory for $0 < g_f \ll 1$. To this end, expand the final temperature in orders of $g_f$,
\begin{equation}
    T_f = T_f^{(0)} + T_f^{(1)} + T_f^{(2)} + \ldots.
\end{equation}
The zeroth order result is readily found as 
\begin{equation}
    T_f^{(0)} = \frac{U}{2 \ell},\quad \ell = \ln \pqty{\frac{U/2}{\abs{E_i / N}}}. 
\end{equation}
As the spin energy comes with a prefactor of $g_f^2$, it suffices to evaluate it at the leading temperature $T_f^{(0)}$. This is much larger than the Heisenberg scale, and we therefore replace $E_\textrm{s}(g_f,T_f) \to E_\textrm{s}(g_f,T \to \infty) \equiv g_f^2 E_{\textrm{s}}^\infty$. Here, we introduced the infinite temperature energy $E_{\textrm{s}}^\infty = E_\textrm{s}(1,T \to \infty) = -N z_\textrm{lat} J/8$, where $z_\textrm{lat}$ is the lattice coordination number. Similarly, we note that at small $x$, $r(x) \simeq x/2$ such that the term $-2d t_f r(2 t_f / T_f) \simeq - 2 d g_f^2 t^2 /T_f^{(0)}$ to leading order. This will only give rise to contributions that are subleading by an additional factor of $(t/U)^2$, so we will henceforth neglect this term here. A similar argument applies for the prefactor $[I_0(2t_f/T_f)]^d$. Solving the resulting equation order by order in $g_f$, we obtain
\begin{equation}
    T_f^{(1)} = - g_f \frac{2}{\ell^2} \frac{U}{2},\quad T_f^{(2)} = g_f^2 \pqty{\frac{4}{\ell^3} - \frac{2 - E_{\textrm{s}}^\infty/E_i}{\ell^2}} \frac{U}{2}.
\end{equation}

Thus, for initial temperature $T_i \sim J$ we find reduced temperatures at weak quenches, whereas for strong quenches the temperature reaches $T_f \sim U \gg J$. To address where the cross-over between these two regimes occurs, we ask at what point the postquench energy $(2g_f - 1)E_i$ can no longer be accommodated by the spin system alone such that physical doublons have to be present in the final state. At this point the temperature should satisfy $J \ll T_f \ll U$, such that the energy balance becomes 
\begin{equation}
    (2g_f - 1)E_i \simeq g_f^2 E_{\textrm{s}}^\infty,
\end{equation}
where we replaced the spin energy by its infinite temperature expression. Noting that under our assumptions it is $E_i / E_{\textrm{s}}^\infty > 1$, a solution with $1/2 < g_f <1$ is given in Eq.~\eqref{eq:crossover_g_f} in the main text.

\end{document}